\documentstyle[preprint,aps]{revtex}
\begin{document}
\draft
\title{Cosmological Rotation of Quantum-Mechanical Origin\\
and Anisotropy of the Microwave Background}
\author{L. P. Grishchuk}
\address{McDonnell Center for the Space Sciences, Physics Department}
\address{Washington University}
\address{St. Louis, Missouri 63130-4899}
\address{and}
\address{Sternberg Astronomical Institute, Moscow University}
\address{119899 Moscow, V-234, Russia}
\maketitle
\begin{abstract}
It is shown that rotational cosmological perturbations can be
generated in the early Universe, similarly to gravitational waves. The
generating mechanism is quantum-mechanical in its nature, and the created
perturbations should now be placed in squeezed vacuum quantum states. The
physical conditions under which the phenomenon can occur are formulated.
The generated perturbations can contribute to the large-angular-scale
anisotropy of the cosmic microwave background radiation. An exact formula
is derived for the angular correlation function of the temperature
variations caused by the quantum-mechanically generated rotational
perturbations. The multipole expansion begins from the dipole component.
The comparison with the case of gravitational waves is made.
\end{abstract}
\vskip 5mm
\newpage
\section{Introduction}

The recent discovery of the large-angular-scale anisotropy of the
cosmic microwave background radiation (CMBR)~[1] is leading
to important implications.  It shows that the deviations of our
Universe from homogeneity and isotropy extend to the very large
linear scales, of order and longer than the
present day Hubble radius $l_H$. The main contribution to the
large-angular-scale anisotropy is usually provided by perturbations
with wavelengths $\lambda$ of order of $l_H$. In other words, the
lower index multipole components of $\delta T/T$, such as quadrupole
$(l=2)$, octupole $(l=3)$, {\it etc.,} are usually dominated by such
long wavelength perturbations. In general, all possible wavelengths
contribute to
every single multipole. One can devise a spectrum of cosmological
perturbations in such a way that the contributions to, say, quadrupole
component will be dominated by perturbations with $\lambda \gg l_H$ or,
instead, by perturbations with $\lambda \ll l_H$. But, if the spectrum is
not exceedingly ``red'' or ``blue'', the quadrupole component will be
dominated by perturbations with wavelenghts $\lambda$ of order of $l_H$.
This means that the existence of the very long wavelength cosmological
perturbations can now be considered as an observational fact. (We
assume, of course, that the measured $\delta T/T$ is a genuine
cosmological effect and not, say, a result of the poor data processing
or unaccounted sources of different nature.) It is very
likely that perturbations with so long wavelengths are ``primordial'',
survived from the epochs when the Universe was much younger.
It is hard to arrange for these perturbations to have been generated
or contaminated by the local physical processes within the present
day Hubble radius. If so, what is the origin and nature of these
perturbations?

Following Lifshitz~[2], it is customary to identify the
perturbations superimposed on homogeneous isotropic cosmological models
(Friedman-Lemaitre-Robertson-Walker, FLRW, models) as density perturbations,
rotational perturbations and gravitational waves. This division is
dictated by the classification of the $3 \times 3$ symmetric matrix
whose elements describe all the independent components of the perturbed
gravitational field.  Two transverse-transverse components correspond
to gravitational waves, two transverse-longitudinal components correspond
to rotational perturbations, and two remaining components, one of which
is scalar and another is longitudinal-longitudinal, correspond to density
perturbations. The associated gravitational field of all of these
perturbations is capable of affecting the propagating photons of CMBR
and producing the angular anisotropy $\delta T/T$~[3]. So, in principle,
one of these perturbations or all of them together could be responsible
for the observed $\delta T/T$. However, in the framework of classical
cosmology, there is not much to say about the expected types, amplitudes
and spectra of the perturbations. In classical cosmology, perturbations
are put in game ``by hand''. If there is no initial perturbations~---~there
is nothing to discuss.

It is remarkable that the situation changes dramatically if one takes
into account the principles of quantum mechanics. It turns out that there
may be no need anymore for initial perturbations, there may be no need for
complicated generating mechanisms. It may be sufficient to have just what we
already have: the variable gravitational field of the nonstationary
homogeneous isotropic universe and the zero-point quantum fluctuations
of cosmological perturbations listed above. The strong variable
gravitational field of the early Universe plays the role of
the ``pump'' field. It can supply energy to the zero-point quantum
fluctuations and amplify them. It was first demonstrated for gravitational
waves~[4], and we will do here the same for rotational perturbations.
In a more strict and precise language, the perturbations are parametrically
coupled to the variable gravitational pump field and their interaction
Hamiltonian is time-dependent. The initial vacuum state of the quantized
perturbations transforms, as a result of the quantum-mechanical Schr\"odinger
evolution, into a multiparticle state known as a squeezed vacuum state.
This applies to gravitational waves and density perturbations~[5],
and to rotational perturbations too, as will be shown below.
The quantum-mechanically generated perturbations can be useful
for many purposes. Among other things, they are capable of producing
$\delta T/T$ (for gravitational waves, see~[6]).

It is important, however, to stress from the very beginning that
there is a significant qualitative difference between gravitational waves
on one side and rotational and density perturbations on the other side.
Gravitational waves exist in empty space, they do not require any material
medium for their definition. One or another sort of matter in FLRW models is
only needed to govern the variable gravitational pump field but has
nothing to do with gravitational waves themselves. Moreover, there is no
ambiguity as for how gravitational waves couple to the pump field. The
form of coupling follows from the Einstein equations and is such
that the waves can be amplified indeed. In contrast, rotational
and density perturbations are only defined if there is a cosmological
medium, they are perturbations in the medium. The solutions to Einstein
equations give zero for all non-transverse-transverse components
of the perturbed gravitational field, if there is no matter.
The difference between gravitational waves and other cosmological
perturbations is about the same as the difference between photons,
which exist in empty space, and phonons and various ``excitons''
which exist only in a condensed matter. It is necessary for
the primeval cosmological medium, which has been filling the Universe
in the very distant past, to be capable of supporting free oscillations
of density and rotation regardless of presence or absence
of the pump field. In other words, a deformed element of the primeval
medium should have been capable of experiencing the necessary restoring
forces even if the element had been experimented with in our laboratory
on Earth. Moreover, the coupling of these ``harmonic oscillators''
to the gravitational pump field should be appropriate if one
is intended to amplify their zero-point quantum fluctuations.
Thus, every attempt to apply the quantum-mechanical generation
mechanism to density and rotational cosmological perturbations
will always require the fulfillment of additional physical hypotheses,
as compared with the case of gravitational waves, and we will not
be able to avoid making such additional hypotheses in what follows.

It is very interesting that the behaviour of particles and fields in
the nonstationary Universe bothered E. Schr\"odinger as long ago as
in 1939~[7].  His paper is remarkable in several aspects.
It is instructive to follow his argumentation in order to understand
better the physics involved and what we will be doing below.

Schr\"odinger considers ``the familiar wave equation of the second order
\begin{equation}
   -\Delta \psi + {1\over c^2} {\partial^2\psi \over \partial t^2}
   + \mu^2\psi = 0
\end{equation}
($\mu =0$ for light, $\mu =2\pi mc/h$ for material particles)'' which
``is to be regarded as the covariant equation''
\begin{equation}
  {1\over \sqrt{-g}} {\partial \over \partial x_\alpha}
  \left( g^{\alpha\beta} \sqrt{-g}
         {\partial\psi \over \partial x_\beta}
  \right) + \mu^2 \psi = 0
\end{equation}
specialized for the line element of the non-static universe
\begin{equation}
    ds^2 = c^2dt^2 - R^2(t)dl^2
\end{equation}
where $dl^2$ corresponds, in his paper, to a 3-sphere. He discusses ``the
decomposition of an arbitrary wave function into proper vibrations'' and
notices that the positive and negative frequencies of proper vibrations
``cannot be rigorously separated in the expanding universe''.
He says: ``Generally speaking this is a phenomenon of outstanding importance.
With particles it would mean production or annihilation of matter, merely
by the expansion, whereas with light there would be a production of
light travelling in the opposite direction, thus a sort of reflection
of light in homogeneous space''. He also describes this as a ``mutual
adulteration of positive and negative frequency terms in the course
of time'' giving rise to what he calls ``the alarming phenomena''.
Schr\"odinger concludes: ``They are certainly very slight, though,
in two cases, {\it viz.} (1)~ when $R$ varies slowly (2)~when it
is a linear function of time'', and he also says, with sign of relief:
``in this case the alarming phenomena do not occur, even within
arbitrarily long periods of time'', ``there is nothing like a
secularly accumulated pair production''.

 From the position of current knowledge, we can make some comments
to the Schr\"odinger's paper. (It is easy to be critical of a great
physicist's paper if you are doing this more than half a century later!)

First, Eq.~(2) with $\mu =0$ is not the equation for light, it is
the equation for a hypothetical ``massless scalar particle''.
As for light, Schr\"odinger would, perhaps, be pleased to know
that electromagnetic waves in the non-static universe (3)~are totally
``immune'', there is no ``mutual adulteration'' and ``reflection''
of electromagnetic waves at all. The covariant Maxwell equations
\[
   {\partial F_{\mu\nu} \over \partial x^\alpha}
  +{\partial F_{\nu\alpha} \over \partial x^\mu}
  +{\partial F_{\alpha\mu} \over \partial x^\nu} = 0 \,
\]
\[
   {F_{\mu\nu;}}^\nu = 0
\]
specialized for the line element of the non-static universe (3) have
solutions $F_{\mu\nu}$ exactly the same as solutions $F_{\mu\nu}$
to the Maxwell equations in the Minkowski world.  The scale factor
$R(t)$ participates in the electromagnetic energy-momentum tensor
and makes the energy density vary as $R^{-4}$, which is consistent,
of course, with photons being red-shifted or blue-shifted.
But, if one sends light at $t=t_1$ and receives at $t=t_2$,
the result is totally independent of the variability rate of
$R(t)$ in the interval of time between $t_1$ and $t_2$
(in sharp contrast to what takes place for gravitational waves).

Second, Schr\"odinger's conclusion about the extreme weaknees of the
``alarming phenomena'' when $R(t)$ is a linear function of time,
for the kind of the wave equation with $\mu =0$ he studies,
is premature. This statement requires us to go into some details.
Schr\"odinger takes the function $\psi$ as a product of the
time-dependent amplitude $f(t)$ and a spatial eigenfunction
with the wave number $n$. For $f(t)$ he derives the equation
\begin{equation}
    R^{-3} {d\over dt} \left( R^3 {df\over dt} \right)
 + {c^2n(n+2)\over R^2} f= 0 \, .
\end{equation}
We may note that the same equation in the spatially flat universe (3) would
read
\begin{equation}
    R^{-3} {d\over dt} \left( R^3 {df\over dt} \right)
 + {c^2 n^2\over R^2} f= 0
\end{equation}
and
\begin{equation}
    u^{\prime\prime} + \left( n^2 -{R^{\prime\prime} \over R} \right)
    u = 0
\end{equation}
if one makes the substitutions
$u=fR$, $R(\eta )d\eta =cdt$, ${}^\prime = d/d\eta$.

Schr\"odinger investigates the special case
\begin{equation}
   R = a+ bt \quad .
\end{equation}
He writes: ``Specialising for light ($\mu = 0$) and putting for the moment
\begin{equation}
    {c^2n(n+2) \over b^2} = k^2 +1
\end{equation}
we have ... $f = {1\over R} R^{\pm ik}$\, ''. He comes to the conclusion
that the solution $f(t)$ is oscillatory, with the amplitude being
proportional to $R^{-1}$, and that the waves travelling in opposite
directions keep rigorously separated for any length of time.
This conclusion is a consequence of his following statement: ``Since
$n$ is very large, $k$ is real, for $b$ is certainly not much larger
than $c$''.

It is here where the core of the problem lies. Presumably, it is
hard for Schr\"odinger to accept that the velocity of change
of the scale factor can be larger than the velocity of light,
this is why: ``$b$ is certainly not much larger than $c$''.
If he relaxed this assumption, the left hand side of Eq.~(8)
could be made less than 1 and, hence, $k$ could be made
imaginary, at least, for some values of $n$.  This could
be done even in the closed universe, where the set of $n$
is discrete and begins from some lowest wave number labelling
the longest wavelength which can still fit into the universe. In the
spatially flat universe, the situation is even better.
Equation~(8) now reads
\begin{equation}
   {c^2n^2 \over b^2} = k^2 + 1
\end{equation}
where $n$ is continuous and can go to zero. In this case,
one can even accept $b < c$, as there will always be some $n$'s
for which $k$ is imaginary.

The factor $(cn/b)$ plays a crucial role. This factor
is the ratio of the  characteristic time-scale on which $R(t)$
varies and the period of wave with the wave number $n$.
According to Schr\"odinger's assumptions this ratio is much
larger than 1. But this is precisely the condition which allows
us to say that ``$R(t)$ varies slowly'', adiabatically!
If this condition is satisfied, the second Schr\"odinger's
case belongs, in fact, to the same category as the first case.
There is no wonder that the effect is ``very slight''.
Schr\"odinger's assumptions exclude the possibility of
the ``alarming phenomena'' from the very beginning.
It is precisely when this ratio is less than 1 that
the character of the solution for $f(t)$ changes,
it ceases to be oscillatory, and the final effect becomes
significant. It is not a particular functional dependence $R(t)$
itself that is important, but comparison of the time-scale of
variation of $R(t)$ with the period of a given wave.

This can be conveniently illustrated with the help of Eq.~(6).
In the interval of time where $R(t)$ obeys Eq.~(7),
the height of the ``potential barrier''
$U(\eta ) \equiv R^{\prime\prime} /R$
is just a constant:  $R^{\prime\prime} /R = b^2/c^2$.
One can arrange for the $R(\eta )$ to be such that $U(\eta )$
would go to zero, more or less gradually, outside this interval
of time. Then, for the high-frequency waves, {\it i.e} for
$n^2 \gg b^2/c^2$, the solution is $u=A\sin (n\eta + \varphi )$,
where $A$ is almost a constant, the ``alarming phenomena''
do not occur. However, for the low-frequency waves, for which
$n^2 \ll b^2/c^2$, and which, therefore, hit the barrier,
the solution is $u=A\sin (n\eta +\varphi_1)$ initially,
that is far to the left of the barrier, and
$u = B\sin (n\eta + \varphi_2)$ finally, that is far to the right
of the barrier. The remarkable fact is that one always gets
${\bar B}^2 \gg A^2$ (the bar denotes averaging over the arbitrary
initial phase $\varphi_1$), and the wave amplification takes place.
It is not even important whether the universe was expanding
or contracting. With the spatial dependence taken into account,
the effect means amplification of the travelling wave and simultaneous
generation of the ``reflected'' wave, exactly as Schr\"odinger envisaged.

Equations (5), (6) are precisely the equations for the time-dependent
amplitudes of gravitational waves (not electromagnetic waves).
They have been investigated in~[4] and the scale factor
of the form~(7) has been considered as an example (unfortunately,
I did not know about the Schr\"odinger's paper at that time
and became aware of it much later). The quantum-mechanical version
of the process considered above leads to the generation of gravitons
(gravitational waves). The initial vacuum state evolves into a strongly
squeezed vacuum state with the expectation number of gravitons much larger
than one. One can also think of this process, in classical terms, as of
generation of standing gravitational waves. It is a pity that we will
never know whether Schr\"odinger would qualify the production of
gravitational waves as an ``outstanding'' or ``alarming'' phenomenon.

It is necessary to say that a systematic study of the quantized
version of Eq.~(2) was first undertaken by L. Parker~[8].
He investigated the general principles of the field quantization
in spatially flat FLRW models and formulated a number of important
theorems. In particular, Parker emphasizes the important role
of conformal invariance and exhibits the ambiguity in the choice
of the equation for scalar particles. He recognizes that the zero
or nonzero production of the massless scalar particles depends on the
chosen form of the wave equation (form of coupling of particles
to the external gravitational field). However, one of his conclusions,
``we show that massless particles of arbitrary nonzero spin,
such as photons or gravitons, are not created by the expansion,
regardless of its form'', turned out to be incorrect in part of gravitons.

The conviction that massless particles can not be created in
FLRW gravitational fields was predominant in late 60th~--~early 70th.
Since the creation of massive particles (with masses less than the
Planckian) was shown to be negligibly small, it has lead to study
of anisotropic cosmologies. It was believed that only in this case
the effects are nonvanishing and interesting. For instance, Zeldovich
and Starobinsky, in an influential paper~[9], have laid a foundation
for quantization of fields in the spatially flat anisotropic models.
One of their conclusions is: ``It is established that vacuum polarization
and production of particles of a zero mass field are absent only in the
isotropic case'', and they call the isotropic case ``degenerate''.

The interest toward quantized fields and quantized cosmological perturbations
has increased in the context of the inflationary hypothesis~[10].
The inflationary models belong to the class of FLRW homogeneous isotropic
models and have a specific form of the scale factor $R(t)$.
The authors of papers on quantized perturbations often associate
the generating mechanism with such things as ``inflation'', ``rolling
the scalar field down the inflationary potential'', ``horizons'', etc.
The reader can find a review of the extensive literature on the subject
in~[11]. The important early references on the quantization of density
perturbations are~[12,13].

Our goal is to study the quantum-mechanical generation of  rotational
perturbations in FLRW cosmological models. The notion of a gravitational
pump field acting on a harmonic oscillator, often used above, may seem
to be too vague and overly mechanistic. In my view, however, this notion
accurately and precisely describes the physics involved, if one is willing
to use the ``field-theoretical'' formulation of Einstein's general
relativity (see, for instance,~[14]). In this approach, the FLRW
cosmological models get represented as gravitational fields in Minkowski
space-time and the scale factor $R(t)$ manifests itself, literally,
as the gravitational pump field. The coupling of the gravitational
pump field to the quantized gravitational perturbations is provided
by the nonlinear character of the total gravitational field.
The form of coupling follows automatically (at least, in case
of gravitational waves) from the Einstein equations which acquire
now the form of nonlinear wave equations in Minkowski space-time. Along
this route, one can easily demonstrate the far-reaching analogy between
the quantum-mechanical generation of cosmological gravitational waves
and the generation of squeezed light in the laboratory settings of quantum
nonlinear optics~[15]. However, the ``field-theoretical'' approach
is not familiar to, or not accepted by, the most of researchers on
the subject, and, therefore, we will be mostly using here the usual
geometrical formulation.

The basic equations for rotational perturbations in the spatially
flat FLRW cosmological models are presented in Sec.~2.
As one of our physical hypotheses, we assume that the perturbations
could sustain as torque oscillations in the primeval cosmological medium.
In other words, we assume that the torsional velocity of sound was not
equal to zero and, in fact, it could be as big as the velocity of light.
We also assume that the coupling of the torque oscillations to the
curvature was ``minimal'', that is the same as for gravitational waves.
Under these conditions, the time-dependent amplitudes of the rotational
perturbations satisfy Eq.~(6) and the most of the results previously
derived for gravitational waves can be used here. The quantization
of rotational perturbations is considered in Sec.~3.  The important
difference with the case of gravitational waves is only in the
form of polarization tensors. All conclusions with regard of squeezing
and standing waves are essentially the same. In Sec.~4, as an application
to realistic cosmological models, we consider the scale factors which are
power-law dependent on the $\eta$--time. They include the scale factors
of the inflationary models.  We assume that the torsional velocity
of sound drops to zero at the beginning of the matter-dominated stage.
The torque oscillations get converted into usual rotational perturbations
governed by the conservation law for the angular momentum.
In Sec.~5, we consider the angular anisotropy of the CMBR produced
by rotational perturbations. Rotational perturbations generate
all multipoles of $\delta T/T$ beginning from the dipole component
$(l=1)$. An exact formula is derived for the expected angular
correlation function of $\delta T/T$ caused by rotational perturbations
of quantum-mechanical origin. The results are compared with those for
gravitational waves. A brief summary is given in Sec.~6.
\newpage
\section{Basic Equations for Rotational Perturbations}

We write the unperturbed metric in the form
\begin{equation}
  ds^2 = a^2(\eta )(d\eta^2 - dl^2 )
\end{equation}
where
\begin{equation}
  dl^2 = d{x^1}^2 + d{x^2}^2 + d{x^3}^2 \, .
\end{equation}
The scale factor $a(\eta )$ is governed by matter with
the energy-momentum tensor
$T_{\mu\nu}$.
The nonvanishing components of the unperturbed
$T_{\mu\nu}$
must have the form
$T_o^o =\epsilon (\eta )$,\\
$T_i^k = -p (\eta )\delta_i^k$.
The perturbed metric can be written as
\begin{equation}
  ds^2 = a^2(\eta )
 (\eta_{\mu\nu} + h_{\mu\nu}) dx^\mu dx^\nu \,
\end{equation}
Without loss of generality, one can choose a synchronous coordinate
system, so that $h_{oo} = 0$, $h_{oi} = 0$.  The nonvanishing components
of $h_{\mu\nu}$ form a symmetric $3 \times 3$ matrix $h_{ij}$.

The construction of rotational perturbations is based on the 3-vector
fields $Q_i(x^1,x^2,x^3)$ with vanishing divergence~[2].
They are eigenfunctions of the Laplace operator in the 3-space (11)
and satisfy the equations
\begin{equation}
   {Q_{i,k,}}^k + n^2Q_i = 0 \, \qquad {Q^i}_{,i} = 0 \, .
\end{equation}
All the manipulations with $Q_i$ are to be performed with the help
of the metric tensor defined by Eq.~(11). Using $Q_i$, one can
construct a symmetric tensor $Q_{ij}$ and an antisymmetric tensor
$W_{ij}$:
\[
     Q_{ij} = -{1\over 2n} (Q_{i,j} + Q_{j,i} ) \, , \qquad
     W_{ij} = -{1\over 2n} (Q_{i,j} - Q_{j,i} ) \, .
\]
In Sec.~2 we need only to know the general properties of rotational
perturbations, which are listed above, but we will also discuss their
explicit functional dependence later on.

With a given eigenfunction $Q_i$ is associated the contribution to
$h_{ij}$ which can be written as
\begin{equation}
   h_{ij} = 2h(\eta )Q_{ij} \, .
\end{equation}
The nonvanishing components of perturbations to the energy-momentum tensor
$T_{\mu\nu}$ may only have the following general form
\begin{equation}
   T_i^o = a^{-2} \omega (\eta )Q_i \, , \qquad
   T_o^i = -a^{-2} \omega (\eta )Q_i \, , \qquad
   T_i^k = 2a^{-2} \chi (\eta )n^{-1}Q_i^k \, . \qquad
\end{equation}

The perturbations in the matter distribution and the accompanying
perturbations in the gravitational field are governed together
by the perturbed Einstein equations. The functions
$h(\eta )$, $\omega (\eta )$, $\chi (\eta )$
are to be determined from the equations
\begin{equation}
  h^{\prime\prime} + 2{a^\prime \over a} h^\prime = 2 \kappa\chi (\eta )n^{-1}
\end{equation}
\begin{equation}
   -{n\over 2} h^\prime = \kappa \omega (\eta )
\end{equation}
where $\kappa =8\pi G/c^4$. A consequence of Eqs.~(16), (17) is the equation
\begin{equation}
  \omega^\prime + 2{a^\prime \over a} \omega + \chi = 0 \, .
\end{equation}
One can also consider Eq.~(16) as a consequence of Eqs.~(17), (18).
(Equations~(16)--(18) are consistent with those in Bardeen's
paper~[16] if one corrects the following misprints:
the factor $k/5$ in Eq.~(A2c) should read
$k$ and the factor $k$  in Eq.~(A5) should read
$k/2$.)

Let us first study Eq.~(18). This equation is the linearized version
of the covariant equations
${T_i^\alpha}_{;\alpha} = 0$
which have the meaning of the differential conservation laws in
curved space-time, that is, in presence of gravitational field.
They have, of course, definite physical meaning even when
the gravitational field can be neglected.

In case of a material medium placed in flat space-time, we write
\begin{equation}
     T_{i,o}^o + T_{i,k}^k = 0 \, .
\end{equation}
The components $T_i^o$ describe fluxes of energy (momenta), the components
$T_i^k$ describe fluxes of momentum (stresses). Taking derivatives of
Eq.~(19), one arrives at the equation
\begin{equation}
   \left( T_{i,j}^o   - T_{j,i}^o \right)_{,o}
 + \left( T_{i,k,j}^k - T_{j,k,i}^k \right) = 0 \, .
\end{equation}
The antisymmetric tensor
$\omega_{ij} = T_{i,j}^o -T_{j,i}^o$
is the rotation tensor defined locally, in every point.
The antisymmetric tensor
$f_{ij} = T_{i,k,j}^k - T_{j,k,i}^k$
describes the torque forces acting on an element of the deformed medium.
In flat space-time, the components of the energy-momentum tensor can
be taken in the same form as in Eq.~(15) assuming that
$a(\eta ) =1$
(see also Eq.~(10)). Then, one gets
$\omega_{ij} = -2n\omega (\eta )W_{ij}$
and
$f_{ij} =-2n\chi (\eta )W_{ij}$.
The function $\omega (\eta )$ plays the role of the angular velocity.
The nonzero $\omega (\eta )$ is a confirmation of the fact that we are
dealing with the rotational motion, as opposed to the potential motion.
(We will not be going here into further
details such as presence, or not, of the vortex lines, etc.) By
substituting
$\omega_{ij}$
and
$f_{ij}$
in Eq.~(20) one derives the equation
\begin{equation}
  \omega^\prime + \chi = 0
\end{equation}
which basically expresses a familiar law: the rate of change of the
angular momentum is equal to the torque. This equation is precisely
the limiting form of Eq.~(18) in the limit of $a^\prime /a \rightarrow 0$.

Not every deformed medium experiences torque forces. But if it does,
the medium can support torque oscillations. They arise if a given
element of the medium is displaced from the equilibrium position
at a small angle $\theta (\eta )$.  The restoring force is usually
proportional to
$\theta (\eta )$:  $\chi (\eta ) \sim b^2 \theta (\eta )$,
where
$b^2 = v_t^2/c^2$
and
$v_t$
is the torsional velocity of sound. One can write
\begin{equation}
  \chi (\eta ) = n^2b^2\theta (\eta )
\end{equation}
Since $\omega (\eta ) =\theta^\prime (\eta )$,
Eq.~(21) takes the form of the equation
for a harmonic oscillator with frequency $nb$:
\begin{equation}
  \theta^{\prime\prime} + n^2b^2\theta = 0 \, .
\end{equation}
Equation (22) provides us with the simplest dispersion law known in
theory of vibrations: frequency of oscillations is linearly proportional
to the wave number. In the limit of $v_t \rightarrow 0$, the frequency
goes to zero and the angular velocity $\omega (\eta )$ takes a constant
value (free rotation).

In cosmological problems, one often considers an ideal fluid with
the energy-momentum tensor
$T_{\mu\nu} = (\epsilon +p)u_\mu u_\nu - pg_{\mu\nu}$.
A deformed element of the ideal fluid does not experience torque forces.
The perturbed components of
$T_\mu^\nu$ are
$T_i^o =(\epsilon+ p)u^o \delta u_i$,
$T_i^k =-\delta p\delta_i^k$,
and the tensor $f_{ij}$ vanishes. This means that the
$\chi (\eta )$ should be put equal to zero. The angular momentum
conservation law in the nonstationary Universe, Eq.~(18), leads to
$\omega (\eta ) =$~const~$a^{-2}$.
In other words, rotation in the expanding Universe filled
with the ideal fluid can only decay. (This was recognized by Zeldovich long
ago~[17].) Accordingly, the function
$h^\prime (\eta )$ accompanying rotational perturbations does also decay,
as follows from Eq.~(17). Moreover, if $\omega (\eta )=0$, Eqs.~(16), (17)
require $h(\eta )= {\rm const}$ which means that there is no
perturbations at all, as
$h_{ij} ={\rm const}\, Q_{ij}$ can be reduced to zero by a simple
transformation of spatial coordinates.

Our main physical hypothesis is that the primeval cosmological medium
could have supported torque oscillations. We assume that Eq.~(22) was valid
in the very early Universe. Then, Eq.~(18) reads
\[
   \theta^{\prime\prime} + 2 {a^\prime \over a}
   \theta^\prime + n^2b^2\theta = 0 \, .
\]
Also, with the help of Eq.~(17), one can write Eq.~(16) in the form
\begin{equation}
   h^{\prime\prime} + 2{a^\prime \over a} h^\prime + n^2b^2h = 0
\end{equation}
and, by introducing $\mu (\eta ) =ah(\eta )$, in the form
\begin{equation}
   \mu^{\prime\prime} + (n^2b^2 - a^{\prime\prime}/a)\mu = 0
\end{equation}
Eqs.~(24), (25) govern the time-dependent amplitude of components
$h_{ij}$ of the perturbed gravitational field accompanying torque
oscillations in the medium. These equations have the same form
as equations for the time-dependent amplitude of gravitational waves.
The only difference is that the constant $b$ is not strictly 1,
as in the case of gravitational waves, but can now lie in the
interval from 0 to 1. We will not hesitate to assume that $b$
can be close to 1 or equal 1. In general, the properties of the
primeval cosmological madium (presently unknown) could require the
relationship (22) to be more complicated and could make the torsional
velocity of sound a function of time. Nevertheless, it is sufficient for our
purposes to ignore these possible complications and to consider the simplest
assumptions formulated above.

Having reduced the problem to the problem of a parametrically excited
oscillator, we can expect that torque oscillations in the primeval
cosmological medium could have been generated quantum-mechanically
(see Sec.~3), very much similarly to gravitational waves.
They would evolve as torque oscillations till the primeval medium
became an ideal fluid. At this time, the torsional velocity of sound
drops to zero and the oscillations convert into free rotational
perturbations. Without having support of restoring torque forces,
they will gradually decay in the course of expansion, as was described
above. However, according to this hypothesis, the amplitude of
rotational perturbations in the post-recombination Universe
is not necessarily zero and might have been determined by laws
of quantum mechanics, not by a voluntary choice of initial conditions.

In the contemporary inflationary scenarios, the favorite model of
the primeval medium is one or another modification of a scalar field.
Scalar fields can support neither torque oscillations, nor rotation.
There is no use in scalar fields or ideal fluids in terms of
quantum-mechanical generation of rotational perturbations.
However, whether or not the inflationary stage of expansion
(if it took place at all) was governed by a scalar field or
by something else, is totally unknown (in this context, see,
for instance,~[18] about the superstring-motivated cosmologies).
One may hope that there will never be shortage in the ``microphysical''
models for the constituents of matter in the very early Universe,
as long as they lead to interesting and verifiable consequences.

It is also worth mentioning that there may be some ambiguity in
coupling of the perturbations to the external gravitational field.
We intend to use the simplest formula (22). However, one could
alternatively suggest
\begin{equation}
  \chi (\eta ) = 2n\theta (\eta )
  \left[ b^2 + {1\over n^2}
        {a^{\prime\prime} \over a} \right] \, . \eqnum{22'}
\end{equation}
Both formulas would agree in flat space-time
$(a(\eta ) ={\rm const})$ but the second one would be a different
generalization to the case of the variable $a(\eta )$.
Equation~(22') would lead to
$\mu^{\prime\prime} + n^2b^2\mu =0$, instead of Eq.~(25).
Like in the case of electromagnetic waves, the oscillations
would not be superadiabatically amplified, regardless of
the rate of variability of $a(\eta )$. (Similar ambiguity is also
involved, but rarely acknowledged, in attempts to generate density
perturbations through the amplification of the scalar field
fluctuations.) It is not clear whether the hypothesis (22')
is fully consistent, it requires a separate study,
but I do not think that it can be simply ignored.
\newpage
\section{Quantization of Rotational Perturbations}

In flat 3-space (11), the vector eigenfunctions
$Q_i(x^1,x^2,x^3)$ can be conveniently written as
$Q_i\equiv q_i e^{i{\bf nx}} + q_i^\ast e^{-i{\bf nx}}$.
The second of Eqs.~(13) requires the complex vector
$q_i$ to be orthogonal to $n^i$: $q_in^i=0$.
There are two independent real unit vectors orthogonal
to the unit vector
\[
   {\bf n}/n
 = (\sin\theta\cos\varphi , \sin\theta\sin\varphi , \cos\theta )
\]
and to each other. We take them in the form:
\begin{eqnarray}
   {\stackrel{1}{q}}_{i} = l_i = && (\sin \varphi ,
                                     -\cos\varphi , 0) \nonumber\\
   {\stackrel{2}{q}}_{i} = m_i = && \pm (\cos\theta\cos\varphi ,
            \cos\theta\sin\varphi , -\sin\theta ) \, .\nonumber
\end{eqnarray}
In the definition of the vector $m_i$, the sign $+$ is valid for
$\theta < \pi/2$ and the sign $-$ is valid for $\theta > \pi /2$.
The vectors $l_i$, $m_i$ are the basis for the construction
of the polarization tensors ${\stackrel{s}{P}}_{ij}({\bf n})$,
$s=1,\, 2$:
\[
 {\stackrel{1}{P}}_{ij}({\bf n}) = {1\over n} (l_in_j+l_jn_i)\, ,\quad
 {\stackrel{2}{P}}_{ij}({\bf n}) = {1\over n} (m_in_j+m_jn_i)\, .
\]
They describe the transverse-longitudinal degrees of freedom since
\[
 {\stackrel{1}{P}}_{ij}{n^i\over n} = l_i \, ,\quad
 {\stackrel{2}{P}}_{ij}{n^j\over n} = m_i \,
\]
and
${\stackrel{s}{P}}_{ij} n^in^j=0$.
Other important relations are:
\[
  {\stackrel{s}{P}}_{ij}\delta^{ij}
= 0 \, , \quad
  {\stackrel{s}{P}}_{ij}(-{\bf n})
= {\stackrel{s}{P}}_{ij}({\bf n}) \, , \quad
  {\stackrel{s}{P}}_{ij}{\stackrel{s}{P}}^{\prime ij}
= 2\delta^{ss^\prime} \, .
\]

In flat space-time, the general expression for the components
$h_{ij}$ can be written as
\[
   h_{ij}(t,{\bf y})
 = C{1 \over (2\pi )^{3/2}} \int_{-\infty}^\infty d^3k
   \sum_{s=1}^2 {\stackrel{s}{P}}_{ij}({\bf k})
   {1 \over\sqrt{2\omega_k}}
   \left[ {\stackrel{s}{a}}_{\bf k}
    e^{-i\omega_kt} e^{i{\bf ky}}
 + {\stackrel{s}{a}}_{\bf k}^\dagger
    e^{i\omega_kt} e^{-i{\bf ky}} \right] \, .
\]
For the classical $h_{ij}$-field, the quantities
${\stackrel{s}{a}}_{\bf k}$, ${\stackrel{s}{a}}_{\bf k}^\dagger$
are complex conjugate numbers. In the quantized version of the field,
${\stackrel{s}{a}}_{\bf k}$
and
${\stackrel{s}{a}}_{\bf k}^\dagger$
are the creation and annihilation operators for a rotation wave
traveling in the direction ${\bf k}$. The frequency of the wave is
$\omega_k = v_t|{\bf k} |$.

As usual, the normalization constant $C$ is determined from the
requirement of having energy
${1\over 2}\hbar\omega_k$ in each ${\bf k}$-mode:
\[
 \langle 0 | \int_{-\infty}^\infty t_{oo}d^3y|0\rangle
= {1\over 2} \hbar \int_{-\infty}^\infty d^3 k \, \omega_k
  \sum_{s=1}^2 \langle 0|
  {\stackrel{s}{a}}_{\bf k} {\stackrel{s}{a}}_{\bf k}^\dagger
+ {\stackrel{s}{a}}_{\bf k}^\dagger
  {\stackrel{s}{a}}_{\bf k} |0\rangle \, .
\]
The component $t_{oo}$ of the energy-momentum tensor
$t_{\mu\nu}$
for the field
$h_{\mu\nu}$
has been defined in~[14].  The symbol
$h_{\mu\nu}$
denotes different objects here and in Ref.~14 but in the
approximation that we are working in they coincide.
By using Eq.~(47) from Ref.~14 specified to
$h_{oo}=0$, $h_{oi}=0$ and $h_{\mu\nu} \eta^{\mu\nu} = 0$,
one obtains
\[
  \kappa t_{oo}
= {1\over 4} {h^{ij}}_{,o} h_{ij,o}
- {1\over 8} h^{ij,o} h_{ij,o}
- {1\over 8} h^{ij,k} h_{ij,k}
+ {1\over 4} h^{ij,k} h_{jk,i} \, .
\]
The second and third terms cancel each other if
$b=1$, which we will assume for simplicity in what follows.
The last term gives the factor $(-1/2)$ contribution of the first term.
(In case of gravitational waves, all terms, except the first one,
vanish.) As a result, we find the constant
$C$:  $C = 4\sqrt {2\pi c}$.
Then, we make the rescaling:
\[
  {\bf k}  = {{\bf n} \over a} \, , \quad
  {\bf y}  = a{\bf x}    \, , \quad
  \omega_k = {cn \over a} \, , \quad
             {\stackrel{s}{a}}_{{\bf k}}(t)
           = a^{3/2}{\stackrel{s}{c}}_{{\bf n}} (\eta ) \, .
\]
This allows us to write the normalized expression for
$h_{ij}(\eta ,{\bf x})$
in curved space-time (10) as
\begin{eqnarray}
    h_{ij}(\eta ,{\bf x})
  = 4(2\pi )^{1/2} {l_{pl}\over a} {1\over (2\pi )^{3/2}}
    \int_{-\infty}^\infty d^3n
    \sum_{s=1}^2 {\stackrel{s}{P}}_{ij} ({\bf n})
    {1\over \sqrt{2n}}
    \left[ {\stackrel{s}{c}}_{\bf n} (\eta )e^{i{\bf nx}}
         + {\stackrel{s}{c}}_{\bf n}^\dagger (\eta )
    e^{-i{\bf nx}} \right] \, .
\end{eqnarray}
In this expression, $a(\eta )$ has the dimension of length,
$l_{pl} = (G\hbar /c^3)^{1/2}$
is the Planck length, all other quantities are dimensionless.

As in case of gravitational waves~[6], the time evolution of the
operators
$c_{\bf n}(\eta )$, $c_{\bf n}^\dagger (\eta )$ (for each s)
is governed by the Heisenberg equations of motion
$dc_{\bf n}/d\eta  = -i[c_{\bf n},H]$,
$dc_{\bf n}^\dagger /d\eta  = -i[c_{\bf n}^\dagger ,H]$
with the Hamiltonian
\[
  H = nc_{\bf n}^\dagger c_{\bf n}
    + nc_{-{\bf n}}^\dagger c_{-{\bf n}}
    + 2\sigma (\eta )c_{\bf n}^\dagger c_{-{\bf n}}^\dagger
    + 2\sigma^\ast (\eta ) c_{\bf n} c_{-{\bf n}}
\]
where $\sigma (\eta )= ia^\prime /2a$.
The solution to the Heisenberg equations of motion is
\[
    c_{\bf n} (\eta )
  = u_n (\eta ) c_{\bf n}  (0)
  + v_n(\eta ) c_{-{\bf n}}^\dagger (0) , \quad
    c_{\bf n}^\dagger (\eta )
  = u_n^\ast (\eta ) c_{\bf n}^\dagger (0)
  + v_n^\ast (\eta ) c_{-{\bf n}} (0)
\]
where  $c_{\bf n}(0)$, $c_{\bf n}^\dagger (0)$
are the initial values of the operators taken at some initial time
$\eta = \eta_0$.
The complex functions
$u_n(\eta )$, $v_n(\eta )$ satisfy the equations
\[
   iu_n^\prime = nu_n + i(a^\prime /a)v_n^\ast \, , \quad
   iv_n^\prime = nv_n + i(a^\prime /a)u_n^\ast
\]
where
$|u_n|^2 -|v_n|~^2 = 1$
and
$u_n(0)=1$, $v_n(0)=0$.
One can easily show that the function
$\mu_n \equiv u_n+v_n^\ast$
should satisfy the equation
$\mu_n^{\prime\prime} + (n^2-a^{\prime\prime} /a)\mu_n=0$
and the initial conditions
\begin{equation}
  \mu (0) = 1 \, , \quad
  \mu^\prime (0) = -in + (a^\prime /a)(0) \, .
\end{equation}
For both polarizations $s=1,2$, the solutions are identically the same.

The operators
${\stackrel{s}{c}}_{\bf n}(0)$, $\stackrel{s}{c}_{\bf n}^\dagger (0)$
obey the commutation relations
$[{\stackrel{s}{c}}_{\bf n} (0),
{\stackrel{s^\prime}{c}}_{\bf m}^\dagger (0)]
=\delta^{{ss^\prime}} \delta^3 ({\bf n}\!-\!{\bf m})$
and the same is true for the evolved operators:
$[{\stackrel{s}{c}}_{\bf n} (\eta ),
{\stackrel{s^\prime}{c}}_{\bf m}^\dagger (\eta )]
=\delta^{{ss^\prime}} \delta^3({\bf n}\!-\!{\bf m})$.

In the Schr\"odinger picture, the initial vacuum state evolves into
a strongly squeezed vacuum state. The squeeze parameters can be derived
from the solutions for $u_n(\eta )$, $v_n(\eta )$ (see~[6] and
references therein). The total linear momentum and the total angular
momentum of the field remain equal to zero. In the Heisenberg picture,
which we will adopt for further calculations, the quantum state
remains vacuum while all the dynamical properties of the field
are described by the time-dependent annihilation and creation operators.
In Sec.~5 we will be calculating the expectation values with respect
to the vacuum state defined by
${\stackrel{s}{c}}_{\bf n}(0)|0\rangle =0$.

As is clear from the physical description of the phenomenon, there
is nothing specifically ``gravitational'' or ``cosmological'',
let alone ``inflationary'', in the amplification process.
The kind of parametric interaction that is provided by
$a(\eta )$ in cosmology can be realized by non-gravitational
means in the laboratory conditions. It is not excluded that
the squeezed vacuum state for the analogue of torsional oscillations
can be generated in the quantum optics experiments similar,
for example, to those discussed in~[15].
\newpage
\section{Application of The Quantum-Mechanical Generating Mechanism
to Simple Cosmological Models}

The presently available observational data about our part of the Universe
refer to the interval of evolution that took place since the epoch of
the primordial nucleosynthesis. The major uncertainties in the functional
form of the scale factor $a(\eta )$ pertain to the earlier times.
This earlier part of evolution was governed by unknown primeval
cosmological medium with unknown equation of state. The ``microphysical''
details of the primeval medium are quite irrelevant as long as we
are only interested in the scale factor $a(\eta )$.  The function
$U(\eta )=a^{\prime\prime}/a$
is essentially all we need to know in order to find solutions to
Eq.~(25) and calculate the present day characteristics of the
quantum-mechanically generated perturbations. The origin
of perturbations makes it clear that it is only $a(\eta )$
(the pump field) that is directly involved and responsible
for the final result. For example, the spectrum of gravitational
waves is in one to one correspondence with the variable Hubble
parameter of the very early Universe and can be used for its
reconstruction~[19].

We will explore the $\eta$-time power-law scale factors:
$a(\eta ) = a_o \eta^{1+\beta}$,
where $\beta$ is a constant. This parameterization has been used
before~[4]. It provides a simple description of the two cases when
the barrier $U(\eta )= a^{\prime\prime} /a$ vanishes: $\beta =-1,\,0$.
The value $\beta =-1$ gives the flat space-time, the value
$\beta =0$ gives the scale factor of the radiation-dominated Universe.
The matter-dominated Universe is described by $\beta =1$.
We call evolution controlled by $\beta =0$, $\beta =1$ the $e$-stage
and $m$-stage, respectively. Any value of $\beta$ smaller than $-1$
describes a sort of inflationary ($i$-stage) expansion.
The expansion is inflationary in the sense that the length scale
equal to the Hubble radius at some early time of expansion can grow
at the consecutive $i$, $e$, $m$ stages up to, at least, the size
of the present day Hubble radius $l_H$. The value  $\beta =-2$
corresponds to the De-Sitter case.

Specifically, we consider expanding models with the following
scale factor (see also~[20]):\\
$i$-stage:
\begin{equation}
   a(\eta ) = l_o | \eta |^{1+\beta} \, \qquad
   \eta \leq \eta_1 \, , \quad \eta_1 < 0
\end{equation}
$e$-stage:
\[
   a(\eta ) = l_oa_e(\eta - \eta_e )\, , \qquad
   \eta_1 \leq \eta \leq \eta_2
\]
$m$-stage:
\[
   a(\eta ) = l_oa_m(\eta - \eta_m )^2 \, , \qquad
   \eta_2 \leq \eta
\]
where $a_e =-(1+\beta )|\eta_1 |^\beta$,
$\eta_e =\beta\eta_1 /(1+\beta )$,
$a_m=a_e/4(\eta_2-\eta_e )$, $\eta_m = -\eta_2+2\eta_e$,
and $l_o$ is a constant with the dimensionality of length.
The scale factor $a(\eta )$ and its first derivative
$a^\prime (\eta )$ are continuous at $\eta = \eta_1$ and
$\eta = \eta_2$.  We will denote the present time by
$\eta = \eta_R$. The scale factor at the
beginning of the $m$-stage $a(\eta_2)$
is related to its present day value $a(\eta_R)$
by the relation $a(\eta_2)/a(\eta_R)\approx 10^{-4}$.
The $i$-stage is governed by ``matter'' with the effective
equation of state $p=q(\beta )\epsilon$
where  $q(\beta )=(1-\beta )/3(1+\beta )$.

The general solution to Eq. (25), for a given mode $n$, is
\begin{equation}
   \mu_n =(nb\eta )^{1/2} \left[ A_1J_{\beta+1/2}(nb\eta )
  + A_2 J_{-\beta-1/2} (nb\eta ) \right] \, .
\end{equation}
In order to work solely with the Bessel functions, as two linearly
independent solutions to Eq.~(25), we exclude (temporarily) the
half-integer $\beta$'s. The form of the solution (29) shows explicitly,
once again, that the most important parameter involved is the
ratio of the frequency of the wave to the frequency of variations
of the scale factor (frequency of the pump). In terms of $\eta$-time,
those frequencies are, respectively, $\omega_w = nb$ and
$\omega_p=2\pi a^\prime /a$. Their ratio, which can be called
the parameter of adiabaticity, is
$nb\eta /2\pi (1+\beta )$. For periodic parametric couplings,
the equality  $\omega_p=2\omega_w$ is the condition of
the parametric resonance.  The solution (29) ceases to oscillate
and looses its adiabatic character when the
$\omega_w/ \omega_p$ falls down to a number
of order 1.  This is when the parametric amplification becomes
significant. If $b \ll 1$, this happens long before the wavelength
of the oscillation, $\lambda =2\pi a/n$, gets comparable with
the length of the Hubble radius,  $l=a^2/a^\prime$.

To simplify calculations and make them identical to the case of
gravitational waves, we will assume that the torsional velocity of sound
at $i$-stage was equal to the velocity of light, $b_i=1$.
We will also assume that the value of $b$ at $e$-stage was not zero
(at least, in one unspecified component of the pre-recombination matter)
and could be close to 1, $b_e \approx 1$. And, finally, we assume that $b$
has fallen down to zero at $m$-stage, $b_m=0$.  We will be working
directly with the functions $h(\eta )$, $h^\prime (\eta )$ where
$\mu(\eta ) = ah(\eta )$. It is
$h^\prime (\eta )$ at the $m$-stage that we will eventually need
for calculations of $\delta T/T$.  Then, the solution to Eq.~(25)
can be written at the three stages as follows:\\
$i$-stage:
\begin{equation}
  h(\eta ) = {1\over a(\eta )} (n\eta )^{1/2}
             \left[ A_1 J_{\beta+1/2} (n\eta )
           + A_2 J_{-\beta-1/2}(n\eta ) \right]
\end{equation}
$e$-stage:
\begin{equation}
  h(\eta ) = {1\over a(\eta )}
  \left[ B_1 e^{-inb_e(\eta -\eta_e)}
       + B_2 e^{+inb_e(\eta -\eta_e)} \right]
\end{equation}
$m$-stage:
\begin{equation}
    h(\eta ) = {1\over a(\eta )}
               \left[ C_1 n^2(\eta -\eta_m)^2
             + {C_2\over n(\eta -\eta_m)} \right]
\end{equation}
\[
  h^\prime (\eta )=-{1\over a(\eta )} {3nC_2\over n^2(\eta -\eta_m)^2 }
\]

The initial conditions (27) taken at $\eta =\eta_o$ lead to
\[
  A_1 = -{i\over\cos\beta\pi} \sqrt{{\pi \over 2}}
  e^{i(n\eta_o+\pi \beta /2)} \, \qquad
  A_2 = iA_1e^{-i\pi \beta}
\]
where $n|\eta_o |\gg 2\pi |1+\beta|$.
It is worth noting that had we assumed $b_i=0$, the solution to
Eq.~(25) at the $i$-stage would have been
\[
  h(\eta ) = A_1+A_2\int a^{-2} d\eta
\]
and the initial conditions $\mu (\eta_o)=1$,
$\mu^\prime (\eta_o) = a^\prime /a(\eta_o)$
would have required $A_2 =0$,\\
$h(\eta )= A_1$.
This is another way of saying that the perturbations could not
have been generated quantum-mechanically in medium with zero torsional
velocity of sound.

The coefficients
$B_1$, $B_2$, $C_1$, $C_2$
are determined by the continuous matching of solutions (30)--(32)
at $\eta = \eta_1$ and  $\eta = \eta_2$.
We are interested in waves that have interacted with the barrier
$U(\eta )$ at $i$-stage. Their wave numbers obey the equation
$n|\eta_1 | \ll 2\pi |1+\beta |$.
For the coefficients $B_1$, $B_2$ with these wave numbers, one derives
\begin{equation}
   B_1 \approx -B_2 \approx {1\over 2}
   e^{in\eta_o} (\beta + 1) \psi (\beta )(n\eta_1 )^\beta \equiv B
\end{equation}
where
$\psi (\beta )= \sqrt {{\pi \over 2}}
e^{i\pi \beta/2}[2^{\beta+1/2}\Gamma (\beta+3/2)\cos\beta\pi ]^{-1}$,
$| \psi (\beta )| = 1$ for $\beta =-2$.
The exact formulae for
$C_1$, $C_2$ are as follows
\begin{eqnarray}
     C_1 =
&&  {1\over 12y_2^2}
    \left[ {\bar B}_1 (\sin b_ey_2 + 2b_ey_2\cos b_2y_2)
     + {\bar B}_2 (\cos b_ey_2 - 2b_ey_2\sin b_ey_2) \right] \nonumber\\
     C_2 =
&&  {4\over 3}y_2
    \left[ {\bar B}_1 (\sin b_ey_2 - b_ey_2\cos b_ey_2)
     + {\bar B}_2 (\cos b_ey_2 + b_ey_2\sin b_ey_2) \right] \nonumber
\end{eqnarray}
where
${\bar B}_1=i(B_2-B_1)$, ${\bar B}_2=B_1+B_2$,
and  $y_2=n(\eta_2-\eta_e)$.
The coefficient $C_1$ represents the constant part of $h(\eta )$
which plays no role in our calculations in Sec.~5 and can, in fact,
be eliminated by a coordinate transformation, but we will discuss
$C_1$ together with $C_2$ for the completeness of our analysis.
The formulae for $C_1$, $C_2$, simplify due to Eq.~(33) and their
approximate expressions can be written as
\begin{eqnarray}
            C_1
&& \approx -{1\over 6} iBy_2^{-2}(\sin b_ey_2
          + 2b_ey_2\cos b_ey_2 )\nonumber\\
            C_2
&& \approx -{8\over 3} iBy_2(\sin b_ey_2
          - b_ey_2\cos b_ey_2 )
\end{eqnarray}

The wave number $n_r =1/b_e(\eta_2-\eta_e)$ defined by the condition
$b_ey_2=1$ separates the functional dependence of
$C_1$, $C_2$, on $n$ into two different regimes. For
$n\ll n_r$ one has
\[
C_1\approx -{1\over 2}iBb_e^2\left( {n_r \over n} \right) \, , \qquad
C_2=-{8\over 9}iB{1\over b_e} \left( {n\over n_r} \right)^4 \, ,
\]
and the dependence of $C_1$, $C_2$   on $n$ is
smooth. For $n \gg n_r$, one has
\[
  C_1\approx-{1\over 3}iBb_e^2
  \left( {n_r \over n} \right) \cos {n\over n_r} \, ,\quad
  C_2\approx {8\over 3}iB{1\over b_e}
  \left( {n\over n_r} \right)^2 \cos {n\over n_r} \, ,
\]
and the dependence of $C_1$, $C_2$
on $n$ is oscillatory. There exists a series of frequencies
$n$ at which both, $h(\eta )$ and $h^\prime (\eta )$, go to zero
for all $\eta > \eta_2$. These frequencies are defined by the requirement
that the factor $\cos (n/n_r)$ vanishes. In the framework of quantum
mechanics, one can say that the $n$-modes with those frequencies
are having been ``desqueezed'', stripped off the energy accumulated at
the $i$-stage by the very late times of their evolution at the
$m$-stage~[21].

The $n$-dependent spectrum of the field
$h^\prime (\eta )$, which we will need in our further calculations,
is smooth for $n\ll n_r$ and is oscillatory for $n\gg n_r$.
For a qualitative description of the spectrum we introduce
$h^\prime (n)=l_{pl}(|\mu_n|/a)^\prime$. Then, we can write
\begin{eqnarray}
    h^\prime (n)
&& \approx {1\over 3} {\l_{pl} \over l_o}
    |\psi(\beta )|b_e^2 \,
    {a^2(\eta_2) \over a^2(\eta )}
    {n^{\beta+3}\over n_r} \, ,\qquad n\ll n_r \nonumber\\
    h^\prime (n)
&& \approx {\l_{pl} \over l_o}
  |\psi(\beta )|b_e^2 \,
  {a^2(\eta_2) \over a^2(\eta )} n_rn^{\beta+1}
  \left| \cos {n\over n_r} \right | \, ,\qquad n\gg n_r \, .
\end{eqnarray}
The absolute values of $h^\prime (n)$ are primarily determined by $b_e$
and by the parameters $l_o$, $\beta$ of the $i$-stage.

In the limiting case $b_e = 0$, the form of solution for
$h(\eta )$ is the limiting form of Eq.~(31):
\[
  h(\eta ) = {1\over a(\eta )}
  \left[ {\stackrel{=}{B}}_1 n(\eta -\eta_e )+ {\bar B}_2 \right]
\]
where ${\stackrel{=}{B}}_1$ and ${\bar B}_2$ are constants. The
$h^\prime (\eta )$ at $\eta =\eta_R$ is not zero,
as would follow from the approximate formulae (35) in the limit
$b_e =0$, but is really very small, its value is determined
by small terms omitted in course of derivation of Eq.~(35).
In case of $b_e =0$, rotation is still being generated at the $i$-stage
but decays since the beginning of the $e$-stage and its small amplitude
now makes it probably useless for astrophysical applications.
\newpage
\section{Angular Anisotropy of the CMBR Caused by Rotational Perturbations}

Our major goal is to derive the angular correlation function for the
anisotropy $\delta T/T$ produced by rotational perturbations of
quantum-mechanical origin. However, we will start from the analysis
of the problem at the classical level.

The solution (32) for rotational perturbations is given in a
synchronous coordinate system. It is convenient, first, to go over to a
comoving coordinate system where, by definition, the components
$T_o^i$ of the energy-momentum tensor $T_{\mu\nu}$ vanish.
This allows us to describe the world line of the observer by simple
equations $x^i=0$. It is assumed that the observer's world line
is one of the world lines of the matter. If the observer has
a peculiar velocity, the observed $\delta T/T$ will have an additional
dipole component which is known how to deal with.
The spatial hypersurfaces of constant time should be retained
the same as in the synchronous coordinate system.
The energy density of matter is constant over these hypersurfaces
and the radiation temperature was everywhere the same at the
time of decoupling of the CMBR. This allows us to define the
emission time of the photons of CMBR as $\eta =\eta_E$ regardless
of the direction of observations.

The transition to the comoving coordinate system
is achieved with the help of a small coordinate transformation
\begin{equation}
  {\bar \eta} = \eta , \quad
  {\bar x}^i = x^i -{C_2\over 8}
  {1\over a(\eta )} (\eta -\eta_m)Q^i
\end{equation}
where ${\bar \eta}$, ${\bar x}^i$, are the coordinates of the comoving system.
The transformed metric tensor is
\begin{equation}
    ds^2 = a^2(\eta )(\eta_{\mu\nu} +{\bar h}_{\mu\nu})dx^\mu dx^\nu
\end{equation}
where
\[
{\bar h}_{oo} = 0 , \quad
{\bar h}_{oi} = {C_2\over 8} {1\over a} Q_i = {\bar g}(\eta )Q_{i}, \quad
{\bar h}_{ij} = 2 \left[ h(\eta ) + {C_2 n(\eta -\eta_m) \over 8a} \right]
Q_{ij} = 2{\bar h}(\eta )Q_{ij}
\]
and
\begin{equation}
  {\bar h}^\prime (\eta ) = -{1\over a(\eta )}
  {3nC_2\over n^2(\eta - \eta_m)^2}
  \left[ 1+ {1\over 24} n^2(\eta -\eta_m)^2 \right] , \quad
  {\bar g}^\prime (\eta ) = -{1\over a(\eta)}
  {nC_2\over 4n(\eta - \eta_m)} .
\end{equation}
After performing the transformation, we do not write the bar over
the coordinates $x^\mu$ which are now supposed to be the comoving
coordinates. The transformed components of the perturbed
energy-momentum tensor are
\begin{equation}
    {\bar T}_o^i = 0 \, , \qquad
    {\bar T}_i^o = a^{-2} \omega Q_i \, , \qquad
    {\bar T}_i^k = 2 a^{-2} \chi n^{-1} Q_i^k \, .
\end{equation}

The transition from Eqs.~(12), (15) to Eqs.~(37), (39) has been done
with the help of a usual coordinate transformation.
I prefer to reserve the notions of gauge transformations and gauge
invariance to the ``field-theoretical'' formulation of general
relativity where they have their genuine meaning: distinct from coordinate
transformations, independent of any approximation scheme, unrelated
to any prescribed form of the participating functions, {\it etc}.
However, if one wishes to use a different name for coordinate
transformations of the type of Eq.~(36), one can say that the transition
from Eqs.~(12), (15) to Eqs.~(37), (39) has been performed
with the help of a ``gauge transformation''.

The calculation of the CMBR temperature variations caused by
the gravitational field of cosmological perturbations was first
undertaken by Sachs and Wolfe~[3]. The authors work in the comoving
coordinate system and derive the formula
\begin{equation}
  {\delta T \over T} = {1\over 2} \int_0^{\eta_R-\eta_E}
  \left( {\partial {\bar h}_{ij} \over \partial\eta}
  e^i e^j - 2{\partial {\bar h}_{0j} \over \partial\eta }
  e^j \right )_{(o)} dy \, .
\end{equation}
where $e^k$ is a unit vector in the direction of observations.
This formula is valid for all types of cosmological perturbations.

The expression under the integral in (40) depends on
${\bar h}^\prime (\eta )$, ${\bar g}^\prime (\eta)$, Eq.~(38).
The spectrum of perturbations
in comoving coordinates is different from the one in synchronous
coordinates. In our case, the approximate expressions for the spectrum
$h^\prime (n)$ in synchronous coordinates are given by
Eq.~(35). To derive the spectrum ${\bar h}^\prime (n)$
in comoving coordinates one should combine Eqs.~(38), (35).
For the spectrum ${\bar h}^\prime (n)$ at the present epoch
$\eta = \eta_R$, one finds:
${\bar h}^\prime (n) \sim n^{\beta +3}$ for $n\ll n_H$,
${\bar h}^\prime (n) \sim n^{\beta +5}$ for $n_H \ll n\ll n_r$,
${\bar h}^\prime (n) \sim n^{\beta +3} |\cos (n/n_r )|$
for $n\gg n_r$.

We will now turn to the calculation of the angular correlation
function for
$\delta T/T$ caused by rotational perturbations of quantum-mechanical
origin. For quantized perturbations, the
$\delta T/T$ becomes a quantum-mechanical operator.
In our case, by using Eqs.~(40), (26), (38), we can write
\[
   {\delta T\over T} (e^k) =
   {1\over \pi} l_{pl} \int_0^{w_1} dw \int_{-\infty}^\infty d^3n
   \sum_{s=1}^2 \left[ {\stackrel{s}{r}}_{\bf n} (\eta_R-w)
   {\stackrel{s}{c}}_{\bf n}(0) e^{in_ke^kw}
   + {\stackrel{s}{r}}_{\bf n}^\ast (\eta_R-w)
   {\stackrel{s}{c}}_{\bf n}^\dagger (0)
   e^{-in_ke^kw}\right]
\]
where $w=\eta_R-\eta$, $x^k =e^k w$, $w_1=\eta_R - \eta_E$ and
\[
  {\stackrel{s}{r}}_{\bf n} (\eta_R-w) =
  {\stackrel{s}{P}}_{ij}({\bf n}) e^ie^j f_n
  - 2i{\stackrel{s}{q}}_{j}({\bf n}) e^j {\phi}_n, \quad
  f_n = {1 \over \sqrt{2n}}{\bar h}_n^\prime, \quad
  {\phi}_n = {1 \over \sqrt{2n}}{\bar g}_n^\prime.
\]

The mean value of $\delta T/T$ is obviously equal to zero:
$\langle 0| \delta T/T |0\rangle = 0$.
The expectation value of the angular correlation function is defined
as
\[
   K = \langle 0|{\delta T \over T}
   (e_1^k) {\delta T\over T} (e_2^k)|0\rangle
\]
where
$e_1^k$ and $e_2^k$
are two different unit vectors
with the angle $\delta$ between them,
$\cos \delta =e_1^i e_2^j \delta_{ij}$.
It is easy to show that
\begin{equation}
    K = {1\over \pi^2} l_{pl}^2
        \int_0^{w_1} dw \int_0^{w_1} d {\bar w}
        \int_{-\infty}^\infty
        d^3n \sum_{s=1}^2{\stackrel{s}{r}}_n (\eta_R-w)
	{\stackrel{s}{r}}_n^\ast (\eta_R- {\bar w}) e^{in_k \zeta^k}
\end{equation}
where $\zeta^k=e_1^k w-e_2^k {\bar w}$. One can also show that
\begin{eqnarray}
&&  \sum_{s=1}^2{\stackrel{s}{r}}_n (\eta_R-w)
    {\stackrel{s}{r}}_n^\ast (\eta_R- {\bar w})
    = f_n(\eta_R-w) f_n^\ast (\eta_R -{\bar w}) R^{11} ({\bf n}) \nonumber\\
&&  + 2i f_n (\eta_R-w){\phi}_n^\ast (\eta_R-{\bar w}) R^{12} ({\bf n})
    - 2i f_n^\ast (\eta_R-{\bar w}){\phi}_n (\eta_R-w) R^{21} ({\bf n})
    \nonumber\\
&&  +4 {\phi}_n(\eta_R-w) {\phi}_n^\ast (\eta_R-{\bar w}) R{^22} ({\bf n})
\end{eqnarray}
where
\begin{eqnarray}
&&  R^{11} = \sum_{s=1}^2 {\stackrel {s}{P}}_{ij} e_1^i e_1^j
    {\stackrel {s}{P}}_{ij} e_2^i e_2^j , \quad
    R^{12} = \sum_{s=1}^2 {\stackrel {s}{P}}_{ij} e_1^i e_1^j
    {\stackrel {s}{q}}_{j} e_2^j ,  \nonumber\\
&&  R^{21} = \sum_{s=1}^2 {\stackrel {s}{P}}_{ij} e_2^i e_2^j
    {\stackrel {s}{q}}_{j} e_1^j , \quad
    R^{22} = \sum_{s=1}^2 {\stackrel {s}{q}}_{i} e_1^i
    {\stackrel {s}{q}}_{j} e_2^j . \nonumber
\end{eqnarray}
	The next step is integration of the functions
$R^{11}, R^{12}, R^{21}, R^{22}$
multiplied by $e^{in_k \zeta^k}$
over the
angular variables $\varphi$, $\theta$. Our intention is to derive
the formula in the most general form, applicable to arbitrary
functions $f_n(\eta_R-w)$, ${\phi}_n(\eta_R-w)$, so we leave the integration
over $w, {\bar w}$ and n to the very end. A lengthy calculation gives the
following results. For $R^{11}$ :
\begin{eqnarray}
&&  \int_0^{2\pi} d\varphi \int_0^\pi\sin\theta \,
    d\theta\cos (n_k \zeta^k) R^{11} ({\bf n}) \nonumber\\
&&  = 16\pi\sqrt{{\pi\over 2}}
    \left\{ (3\cos^2\delta -1)(n\zeta )^{-3/2} J_{3/2}(n\zeta ) \right{}
    + [\cos\delta (\cos^2\delta -1)nw \, n {\bar w} \nonumber\\
&& -4 (3\cos^2\delta -1) ] (n\zeta )^{-5/2} J_{5/2} (n\zeta )
  -8\cos\delta (\cos^2\delta -1)
   nw \, n {\bar w} (n\zeta )^{-7/2} J_{7/2}(n\zeta ) \nonumber\\
&& -(\cos^2\delta -1)^2 (nw)^2 (n {\bar w} )^2
   \left{} (n\zeta )^{-9/2} J_{9/2} (n\zeta ) \right\} \, , \nonumber
\end{eqnarray}
for $R^{12}$ :
\begin{eqnarray}
&&  \int_0^{2\pi} d\varphi \int_0^\pi\sin\theta \,
    d\theta\sin (n_k \zeta^k) R^{12} ({\bf n}) \nonumber\\
&&  = 8\pi\sqrt{{\pi\over 2}}
    \left\{ [2 nw \cos\delta - n{\bar w}(3\cos^2\delta -1)] \right{}
    (n\zeta )^{-5/2} J_{5/2} (n\zeta ) \nonumber\\
&&  + (\cos^2\delta -1) nw\, n{\bar w} [ nw- n{\bar w} \cos\delta ]
    \left{} (n\zeta )^{-7/2} J_{7/2}(n\zeta ) \right\} \, , \nonumber
\end{eqnarray}
and for $R^{22}$ :
\begin{eqnarray}
&&  \int_0^{2\pi} d\varphi \int_0^\pi\sin\theta \,
    d\theta\cos (n_k \zeta^k) R^{22} ({\bf n}) \nonumber\\
&&  = 4\pi\sqrt{{\pi\over 2}}
    \left\{2 \cos\delta (n\zeta )^{-3/2} J_{3/2}(n\zeta ) \right{}
    + (\cos^2\delta -1)nw \, n{\bar w}
    \left{} (n\zeta )^{-5/2} J_{5/2} (n\zeta ) \right\} \, , \nonumber
\end{eqnarray}
where $\zeta = (w^2 -2w {\bar w} \cos\delta + {\bar w}^2)^{1/2}$.
The integration of $R^{21} ({\bf n})$ gives the result which differs from
the above one for $R^{12} ({\bf n})$  by the replacement
$w \leftarrow\rightarrow {\bar w}$ and by the
opposite total sign.

One should now use the ``summation theorem''~[23] and
the relations between the Gegenbauer polynomials and
the associated Legendre polynomials [24] which can be
combined together into the formula (for half-integer $\nu$):
\begin{equation}
  (n\zeta )^{-\nu} J_\nu (n\zeta )
= \sqrt{2\pi} \sum_{k=0}^\infty (\nu + k)
  {J_{\nu +k}(nw) \over (nw)^\nu}
  {J_{\nu +k}(n {\bar w} ) \over (n {\bar w})^\nu}
  {d^{\nu-1/2} \over dx^{\nu -1/2}}
  P_{k+ \nu -1/2} (x)
\end{equation}
where $x = \cos\delta$ and $P_l(x)$ are the Legendre polynomials.

In agreement with Eq.~(42), the expression for $K$ can be presented in the
form
\[
   K = K^1 + K^2 + K^3
\]
where the $K^1$ involves products $f_n f_n^ \ast$, the $K^2$ involves products
$f_n {\phi}_n^ \ast$ and $f_n^ \ast {\phi}_n$, and the $K^3$ involves products
${\phi}_n {\phi}_n^ \ast$. By using Eq.~(43) one obtains for $K^1$:
\[
   K^1 = l_{pl}^2 \sum_{l=1}^\infty K_l^1 P_l (x)
\]
where
\begin{equation}
       K_l^1 = {8l(l+1)\over 2l+1}
       \int_0^\infty n^2 \left| \int_0^{w_1}
       {dw\over (nw)^{3/2}}
\left[ (l-1)J_{l-1/2} (nw) -(l+2)J_{l+3/2} (nw) \right]
       f_n(\eta_R -w) \right|^2 dn
\end{equation}
The expression for $K^2$ is more complicated:
\begin{eqnarray}
&&        K^2 = -16 l_{pl}^2
        \int_0^{w_1} dw \int_0^{w_1} d {\bar w}
        \int_0^\infty n^2 dn
        \left\{\sum_{k=0}^\infty (5/2 + k) \right{}
        {J_{5/2 +k}(nw) \over (nw)^{5/2}}
        {J_{5/2 +k}(n {\bar w} ) \over (n {\bar w})^{5/2}}
        {d^2 \over dx^2} P_{k+2}  \nonumber\\
&&      \times [f_n (\eta_R-w){\phi}_n^\ast (\eta_R - {\bar w})
        (nwx + n{\bar w}(1-3x^2))  \nonumber\\
&&      + f_n(\eta_R - {\bar w}){\phi}_n^\ast(\eta_R -w)
        (n{\bar w}x + nw(1-3x^2)) ] \nonumber\\
&&      -(1-x^2) \sum_{k=0}^\infty (7/2 + k) {J_{7/2 +k}(nw) \over (nw)^{5/2}}
        {J_{7/2 +k}(n {\bar w} ) \over (n {\bar w})^{5/2}}
        {d^3 \over dx^3} P_{k+3} \nonumber\\
&&      \times [ f_n (\eta_R-w){\phi}_n^\ast (\eta_R - {\bar w})
        ( nw - n{\bar w}x )
        + f_n (\eta_R - {\bar w}) {\phi}_n^\ast(\eta_R -w)
        \left{} (n{\bar w} - nwx) ] \right\} \, .
\end{eqnarray}
Finally, the expression for $K^3$ reads as
\[
   K^3 = l_{pl}^2 \sum_{l=1}^\infty K_l^3 P_l (x)
\]
where
\begin{equation}
       K_l^3 = 8(2l+1)l(l+1)
       \int_0^\infty n^2 \left| \int_0^{w_1}
       {dw\over (nw)^{3/2}} J_{l+1/2} (nw)
       {\phi}_n (\eta_R -w) \right|^2 dn
\end{equation}

The total angular correlation function $K$ is rotationally symmetric
(depends only on the angle $\delta$ between the directions of observations)
and its multipole expansion begins from the dipole term $(l=1)$.
The numerical values of the multipole components are different for
different cosmological models. The free parameters $b_e$, $l_o$, $\beta$
of the models considered in Sec.~4 can be chosen in such a way
that the level of the predicted variations would be consistent
with what is actually observed by COBE~[1]. The derivation of
the detailed multipole distributions following
from Eqs.~(44), (45), (46) and the construction of the resulting correlation
function as a function of the separation angle $\delta$ require numerical
calculations. This will be a subject of a further discussion.
\newpage
\section{Conclusions}

We have shown that rotational cosmological perturbations with a very
broad spectrum might have been generated quantum-mechanically in
the very early Universe. We have formulated conditions under which
the phenomenon could take place. The main emphasis has been on long
wavelength perturbations which are probably responsible for
the observed large-angular-scale anisotropy of CMBR. The angular
correlation function was derived, and it was shown that the multipole
expansion begins from the dipole term.  (In the limit of the wavelengths
exceeding the present day Hubble radius the dipole component is
suppressed~[22].)  The numerical values of the expected variations
$\delta T/T$ depend on the parameters of the cosmological models
(essentially, parameters of the cosmological pump ``machine'').
In principle, rotational perturbations alone could account for
the observed anisotropy. The comparison with the case of gravitational
waves~[20] shows, however, that the contribution of gravitational
waves to the large-angular-scale anisotropy is likely to be much larger
than that of rotational perturbations since gravitational waves do not
decay in course of time as fast as free rotational perturbations do.
Moreover, quantum-mechanical generation of gravitational waves
does not require any additional physical hypotheses to be fulfilled,
while the rotational perturbations (and density perturbations) do.
However, it is important to realize that the ``primordial'' cosmological
rotation, although small, could have been generated quantum-mechanically.
The role of the ``primordial'' rotation at smaller linear scales was out
of the scope of the present paper.  Nevertheless, one may speculate that,
if the generating mechanism did really work, the ``seeds'' of rotation
of quantum-mechanical origin may have also played a role at the smaller
linear scales characteristic of galaxies and their clusters (for instance,
it is hard to avoid a question whether the flat rotation curves of spiral
galaxies are an evidence for dark matter, or a remnant of the "primordial"
rotation).
\vskip 5mm

This work was supported in part by NASA Grant No.~NAGW~2902 and
NSF~Grant No.~89-22140.

\end{document}